\documentclass[twocolumn,prstab,floatfix]{revtex4}
\usepackage{graphicx}
\usepackage{amsmath}
\usepackage{latexsym}
\usepackage{amsfonts}
\usepackage{amssymb}

\newcommand{\mat}{\bf}
\newcommand{\roq}{\left(\frac{R}{Q}\right)}

\begin{document}

\title{Recirculating BBU thresholds for polarized HOMs with optical
coupling}
\author{Georg H.~Hoffstaetter}
\author{Ivan V.~Bazarov}
\author{Changsheng Song}
\affiliation{Laboratory for Elementary Particle Physics,
Cornell University, Ithaca, New York 14853}

\begin{abstract}
Here we will derive the general theory of the beam-breakup instability
in recirculating linear accelerators with coupled beam optics and with
polarized higher order dipole modes. The bunches do not have to be at
the same RF phase during each recirculation turn.  This is important
for the description of energy recovery linacs (ERLs) where beam
currents become very large and coupled optics are used on purpose to
increase the threshold current. This theory can be used for the
analysis of phase errors of recirculated bunches, and of errors in the
optical coupling arrangement.  It is shown how the threshold current
for a given linac can be computed and a remarkable agreement with
tracking data is demonstrated.  The general formulas are then analyzed
for several analytically solvable cases, which show: (a) Why different
higher order modes (HOM) in one cavity can couple and cannot be
considered individually, even when their frequencies are separated by
much more than the resonance widths of the HOMs. For the Cornell ERL
as an example, it is noted that optimum advantage of coupled optics is
taken when the cavities are designed with an $x$-$y$ HOM frequency
splitting of above 50MHz. The threshold current is then far above the
design current of this accelerator. (b) How the $x$-$y$ coupling in
the particle optics determines when modes can be considered
separately. (c) That the increase of the threshold current obtainable
by coupled optics and polarized modes diminishes roughly with the
square root of the HOMs' quality factors. Therefore the largest
advantages are achieved with cavities that are not specifically
designed to minimize these quality factors, e.g. by means of HOM
absorbers. (d) How multiple-turn recirculation interferes with the
threshold improvements obtained with a coupled optics. Furthermore,
the orbit deviations produced by cavity misalignments are also
generalized to coupled optics.  It is shown that the BBU instability
always occurs before the orbit excursion becomes very large.
\end{abstract}

\maketitle

\section{Introduction}
In several applications of linear accelerators the charged particle
beam passes through the accelerating structures more than once after
being lead back to the entrance of the linac by a return loop. By this
method the linac can either add energy to electrons several times, or
it can recapture the energy of high energy electrons after they have
already been used for experiments. The former technique is referred to
as recirculating linac, the latter as energy recovery linac (ERL)
\cite{Tigner65_01}.

ERLs have received attention
in recent years since they have the
potential to accelerator currents much larger than those of
non-recovering linacs, and since they have the potential for providing
emittances smaller than those in x-ray storage rings at similar
energies and for similar beam currents. This is due to the fact that
the emittances in an ERL can be as small as that of the electron
source, if emittance increase during acceleration can be avoided.

Several laboratories have proposed high power ERLs for different
purposes.  Designs for light production with different parameter sets
and various applications are being worked on by Cornell University
\cite{CHESS01_03,ERL03_12}, Daresbury \cite{Pool03_01}, TJNAF
\cite{Benson01_01}, JAERI \cite{Sawamura03_02}, Novosibirsk
\cite{Kulipanov98_01}, and KEK \cite{Suwada02_01}. TJNAF has
incorporated an ERL in its design of an electron--ion collider (EIC)
\cite{Merminga02_01} for medium energy physics, while BNL is working
on an ERL--based electron cooler \cite{Benzvi03_01} for the ions in
the relativistic ion collider (RHIC). The work at TJNAF, JAERI and
Novosibirsk is based on existing ERLs of relatively small scale.  The
first international ERL workshop with over 150 participants in early
2005 has also shown the large interest in ERLs that is prevalent in
the accelerator community.

One important limitation to the current that can be accelerated in
ERLs or recirculating linear accelerators in general is the
regenerative beam-breakup (BBU) instability.  The size and cost of all
these new accelerators certainly requires a very detailed
understanding of this limitation. In \cite{hoff04} we have described
this theory for particle motion in one degree of freedom. Here we
generalize this theory to two degrees of freedom, i.e. to accelerators
with polarized HOMs and $x$-$y$ coupling of the particle optics.

For one degree of freedom, a theory of BBU instability in
recirculating linacs, where the energy is not recovered but added in
each pass through the linac, was presented in \cite{Bisognano87_01}.
This original theory was additionally restricted to scenarios where
the bunches of the different turns are in the linac at about the same
accelerating RF phase, such as in the so-called continuous wave (CW)
operation where every bucket is filled.  Tracking simulations
\cite{Krafft87_01} compared well with this theory. This theory
determines above what threshold current $I_{\rm th}$ the transverse
bunch position $x$ displays undamped oscillations in the presence of a
higher order mode (HOM) with frequency $\omega_\lambda$.  If there is
only one higher order mode and one recirculation turn with a
recirculation time $t_r$ in the linac, the following formula is
obtained for $T_{12}\sin\omega_\lambda t_r<0$:
\begin{equation}
I_{\rm th} = -\frac{2c^2}{e(\frac{R}{Q})_\lambda Q_\lambda
\omega_\lambda}\frac{1}{T_{12}^*\sin\omega_\lambda t_r}\ ,
\label{eq:simpbbu}
\end{equation}
where $c$ is the speed of light, $e$ is the elementary charge,
$(R/Q)_\lambda Q_\lambda$ is the impedance (in units of $\Omega$) of
the higher order mode driving the instability, $Q_\lambda$ is its
quality factor. In the case of one degree of freedom,
$T_{12}^*=T_{12}$ is the element of the transport matrix that relates
initial transverse momentum $p_x$ before and $x$ after the
recirculation loop.  A corresponding formula had already been
presented in \cite{Randbook}. Occasionally, additional factors are
found when this equation is stated
\cite{Sereno94,Beard03_01,Merminga01_01}. But in \cite{hoff04} it has
been shown that no such additional factors are required.

The beam transport element $T_{12}$ appears since a HOM produces a
transverse momentum $p_x$ during the first pass of a particle. This
produces a transverse position of $x=T_{12}p_x$ when the particle
traverses the HOM for a second time, and this in turn excites the HOM
itself by means of the wake field $W x = W T_{12}p_x $.

When the mode is polarized with an angle $\theta$, the kick produced
during the first turn corresponds to the momentum $(p_x,p_y) =
p(\cos\theta,\sin\theta)$.  With a coupled optics, the resulting orbit
displacement when the particle reaches the HOM after the return loop
is $\vec x = p(T_{12}\cos\theta + T_{14}\sin\theta , T_{32}\cos\theta
+ T_{34}\sin\theta)$. This excites the higher order mode by
the projection of this displacement onto the wake field, $\vec W \cdot \vec
x = W p (T_{12}\cos^2\theta+(T_{14}+T_{32})\sin\theta\cos\theta +
T_{34}\sin^2\theta)$.

The HOM therefore produces a transverse kick that feeds back to
itself, exactly as in the case with one degree of freedom, only that
$T_{12}$ needs to be replaced \cite{Pozdeyev05} by
\begin{equation}
T_{12}^* = T_{12}\cos^2\theta+(T_{14}+T_{32})\sin\theta\cos\theta +
T_{34}\sin^2\theta\ .
\end{equation}

While equation \ref{eq:simpbbu} is derived with one HOM, for one
degree of freedom it is often a good approximation even when the
cavity has several higher order modes. It was shown in \cite{hoff04}
that different HOMs can be treated individually when their frequencies
differ by more than about $\frac{\omega_\lambda}{2Q_\lambda}$. This
statement does not hold for two degrees of freedom as will be shown in
this paper. Modes cannot in general be treated independently, even
when their frequencies are separated by much more than the width of
the HOM's resonance.

An optics configuration that makes $T_{12}^*$ close to zero in order
to make the threshold current very large has been proposed
\cite{Rand80,Tennant05} and tested for coupled beam transport and
polarized HOMs.  This is a good technique when there is one dominant
HOM. When there are several modes this approach does not apply
directly, even if these modes are separated by more than the width of
their resonance.

The paper is arranged as following: first, a dispersion relation for the
current $I_0(\omega)$ is derived including coupling in a one turn
recirculating linac with one cavity having multiple polarized HOMs. The
smallest real value of $I_0$ that can be obtained with real $\omega$
determines the threshold current. Analytical solutions are given for
the case of two polarized HOMs in one cavity. It is explained how the
dispersion relation for this simple case can be solved
efficiently on a computer, and comparisons to analytical
approximations are presented. Approximations are then given for $N$
polarized modes in one cavity. Subsequently a dispersion relation for
multiple cavities and multiple recirculation loops is derived that can
only be solved numerically with similarly efficient techniques.
Finally, misalignments of cavities are considered to investigate when
these misalignments lead to a very large static displacement of the
beam orbit.

\section{$N$ polarized modes in one cavity}
For simplicity we are here investigating one cavity with $N$
higher-order dipole modes (HOMs), each having a polarization angle
$\theta_\lambda$ to the horizontal. Note that slightly polarized
cavities often have at least two HOMs with similar characteristics,
and looking at a single polarized HOM can therefore be misleading.

The unit vector in the direction of the polarization is $\vec
e_\lambda=\cos\theta_\lambda\vec e_x+\sin\theta_\lambda\vec e_y$. The
effective transverse voltage in each HOM is $V_\lambda$, so that a
particle traversing this HOM obtains a transverse-momentum change of
$\frac{e}{c}\vec e_\lambda V_\lambda$. When $\vec V$ is the vector of
all these $N$ voltages, then the momentum change is
\begin{equation}
\Delta\vec p = \frac{e}{c}{\mat E}\vec V\ ,\ \
{\mat E} = \left(\begin{array}{ccc}\vec e_1\ldots\vec e_N\end{array}\right)\ .
\end{equation}
When the particle returns to the cavity after the return
time $t_r$, the particle's position has changed by
\begin{equation}
\Delta\vec x(t') = {\mat T}\Delta\vec p(t'-t_r)\ ,\ \
{\mat T} = \left(\begin{array}{cc}T_{12}&T_{14}\\T_{32}&T_{34}
\end{array}\right)\ .
\end{equation}
This position will increase the voltage in $V_\lambda(t)$ by the
projection of the mode's wake field $W_\lambda(t-t')\vec e_\lambda$
onto $\Delta\vec x(t')$ times the charge $I(t')dt'$ that excites the
field, i.e. $I(t')W_\lambda(t-t')\vec e_\lambda\cdot\Delta\vec
x(t')dt'$.  Integrating over all contributions to the HOM potentials
leads to
\begin{equation}
\vec V(t)=\frac{e}{c}\int_{-\infty}^t I(t'){\mat W}(t-t'){\mat D}\vec
V(t'-t_r)dt'
\end{equation}
with
\begin{equation}
{\mat W} = {\rm diag}(W_\lambda)\ ,\ \  {\mat D}
=
{\mat E}^T\,{\mat T}\,{\mat E}
\end{equation}
and therefore
\begin{eqnarray}
D_{\mu\mu}
&=&
T_{12}\cos^2\theta_\mu + T_{34}\sin^2\theta_\mu +
\frac{T_{32}+T_{14}}{2}\sin 2\theta_\mu \ ,\nonumber\\
D_{\mu\nu}
&=&
T_{12}\cos\theta_\mu\cos\theta_\nu + T_{34}\sin\theta_\mu\sin\theta_\nu\nonumber\\
&+&
T_{32}\sin\theta_\mu\cos\theta_\nu+T_{14}\cos\theta_\mu\sin\theta_\nu\ .
\label{eq:d21}
\end{eqnarray}
Here $I(t')$ is the current of the bunches that have already traveled
for one turn; in the approximation of short bunches it is given by
\begin{equation}
I(t)=I_0 t_b\sum_{m=-\infty}^\infty \delta(t-t_r-mt_b)\ .
\end{equation}
This transforms the integral equation into
\begin{equation}
\vec V(t)
=
I_0\frac{e}{c}t_b
\sum_{m=-\infty}^\infty {\mat W}(t-t_r-mt_b){\mat D}\vec V(m t_b)
\end{equation}
where $W_\lambda(t)=0$ for $t\le 0$.

The Laplace transform of $\vec V(t)$ can be written as
\begin{equation}
\vec V(t+\delta t_b) = \frac{1}{2\pi} \int_{-\infty-ic_0}^{\infty-ic_0}
\vec{\tilde V}_\delta(\omega') e^{-i\omega' t} d\omega'\ .
\end{equation}
And with the following definition
\begin{equation}
\vec{\tilde V}^\Sigma_\delta(\omega) = \sum_{n=-\infty}^\infty
\vec{\tilde V}_\delta(\omega+\frac{2\pi}{t_b}n)
\end{equation}
one obtains
\begin{equation}
\vec{\tilde V}^\Sigma_\delta(\omega) = t_b\sum_{n=-\infty}^{\infty}
\vec V([n+\delta]t_b) e^{i\omega n t_b}\ .
\end{equation}

Since $\vec{\tilde V}^\Sigma_\delta(\omega)$ is periodic with
$2\pi/t_b$, it has a Fourier series, and its Fourier coefficients are
$\vec V([n+\delta]t_b)$, which shows that $\vec{\tilde
V}^\Sigma_\delta(\omega)$ does not vanish. The transverse motion is
stable when $\vec{\tilde V}(\omega)$ is zero for all $\omega$ with
positive imaginary part. If the current is increased the motion can
become unstable at which point $\vec{\tilde V}(\omega)$ is non-zero
for at least one $\omega$ with positive imaginary part. At threshold
it is therefore non-zero for a real value of $\omega$.

As in \cite{hoff04} we will use $t_r=(n_r-\delta)t_b$ to allow for all
recirculating phases, e.g. $\delta=\frac{1}{2}$ for an ERL. The
integral equation now leads to a relation for these coefficients,
\begin{eqnarray}
&&I_0^{-1}\vec V^\Sigma_0(\omega)\\
&=&
\frac{e}{c}t_b^2 
\sum_{m,n=-\infty}^\infty{\mat W}([n-n_r-m+\delta]t_b){\mat D}
\vec V(m t_b)e^{i\omega nt_b}\nonumber\\
&=&
\frac{e}{c}e^{i\omega n_rt_b}
{\mat W}^\Sigma_\delta(\omega){\mat D}
\vec V^\Sigma_0(\omega)\ .\nonumber
\label{eq:eigen}
\end{eqnarray}
This formulation shows that $\vec V^\Sigma_0(\omega)$ is an
eigenvector of the matrix on the right-hand side, and the
corresponding eigenvalue is $1/I_0$.  Its solution is therefore very
similar to the matrix theory for BBU computation without coupling in
\cite{Bisognano87_01}.  The threshold current can thus be determined
by finding the largest real eigenvalue of this matrix for any real
$\omega$. Due to the symmetry properties ${\mat
W}^\Sigma_\delta(\omega+\frac{2\pi}{t_b}) = {\mat
W}^\Sigma_\delta(\omega)$ and ${\mat W}^\Sigma_\delta(-\omega) = {\mat
W}^{\Sigma *}_\delta(\omega)$, it is sufficient to investigate
$\omega\in[0,\pi/t_b]$ to find the BBU threshold current,
\begin{equation}
I_{th}^{-1} = \frac{e}{c}\max_{\omega\in[0,\pi/t_b]}\{ \Lambda|
e^{i\omega n_rt_b}{\mat W}^\Sigma_\delta(\omega){\mat D}\vec V=\Lambda\vec
V,\Lambda\in I\!\! R\}\ .
\end{equation}

\subsection{Two polarized HOMs in one cavity}
For a large number of HOMs this equation should be
solved numerically, but for two HOMs a analytical
solution is simple. The characteristic polynomial for the eigenvalue $1/I_0$
becomes
\begin{equation}
\left|\left(\begin{array}{cc}
\tilde I^{-1}-D_{11}w_1 &
-D_{12}w_1 \\
-D_{21}w_2 &
\tilde I^{-1}-D_{22}w_2
\end{array}\right)\right|=0\ ,
\label{eq:eigen2}
\end{equation}
with $\tilde I=\frac{et_b}{c}e^{i\omega t_r}I_0$ and $w_\lambda =
\frac{1}{t_b}e^{i\omega\delta t_b}({\mat
W}^\Sigma_\delta)_{\lambda\lambda}$.  Solving this quadratic equation
leads to
\begin{eqnarray}
&&\frac{c}{et_b}e^{-i\omega t_r}I_0^{-1}
=
\frac{D_{11}w_1+D_{22}w_2}{2}\label{eq:sqrt}\\
&&\pm\sqrt{
(\frac{D_{11}w_1-D_{22}w_2}{2})^2
+ w_1w_2 D_{12}D_{21}}\ .
\nonumber
\end{eqnarray}
To reduce the threshold current, the left hand side of this formula
should be small.The matrix ${\mat D}$ is determined by the mode
polarization and by the linear particle optics. It has been suggested
\cite{Rand80,Pozdeyev05} to use these parameters to increase the
threshold current. These suggestions amount to reducing the right hand
side of Eq.~(\ref{eq:sqrt}) by making $D_{11}$ and $D_{22}$ small.
This is always a valid strategy if there is only one HOM,
i.e. $w_2=0$. Then
\begin{equation}
D_{11} = T_{12}\cos^2\theta_1+T_{34}\sin^2\theta_1 +
\frac{T_{14}+T_{32}}{2}\sin 2\theta_1
\end{equation}
and a horizontally ($\theta_1=0$) or a vertically
($\theta_1=\frac{\pi}{2}$) polarized mode together with a beam
transport that fully couples the vertical to the horizontal motion and
vice versa, i.e.~$T_{12}=T_{34}=0$, would always lead to $D_{11}=0$ so
that there would not be any threshold current.  A formula for this
case has been derived in \cite{Pozdeyev05}.

For the case of two or more HOMs this method is no longer as
effective, even when the modes have very different frequencies. In
fact it has been suggested that in cases with several modes, each mode
could be considered separately when each cavity mode has a resonance
width which is significantly smaller than the frequency separation
between modes \cite{Pozdeyev05}. This is however not correct and it
will be seen shortly that the described method that would seem to
increase the threshold current when all modes are considered
separately is not as effective even when modes frequencies differ by
much more than the resonance width. This is especially important when
there are two HOMs of similar properties, for example in nearly
cylindrically symmetric cavities. To see this effect, we distinguish
three cases.

\subsubsection{Circular symmetry}
For circular symmetric cavities there are two equivalent
modes with perpendicular polarization, $w_1=w_2$, $\vec e_1=\vec e_x$,
$\vec e_2=\vec e_y$ leading to ${\mat D}={\mat T}$,
\begin{eqnarray}
I_0 &=& \frac{e}{ct_b}
\frac{e^{-i(\omega t_r+\vartheta_\pm)}}{T_\pm w_1}\ ,
\label{eq:casea}\\
T_\pm e^{i\vartheta_\pm}
&=&
\frac{T_{12}+T_{34}}{2} \pm
\sqrt{(\frac{T_{12}-T_{34}}{2})^2 + T_{14}T_{32}}\ ,\nonumber\\
&& {\rm with}\ T_\pm \in I\!\! R^+\ .\nonumber
\end{eqnarray}
where the threshold current is the smaller of the two values obtained
with the equation for $+$ and for $-$. For $T_{12}=T_{34}=0$ this case
has been considered in \cite{Yunn05}. If there is no coupling,
$T_+=|T_{12}|$ and $T_- = |T_{34}|$.  This result is equivalent to
what has been found in \cite{hoff04} for motion in one degree of
freedom.

As in \cite{hoff04} we now use long range wake fields of the form
\begin{equation}
W_\lambda(\tau)
=
\roq_\lambda\frac{\omega_\lambda^2}{2c}
e^{-\frac{\omega_\lambda \tau}{2Q_\lambda}}\sin(\omega_\lambda\tau)\ ,
\end{equation}
where $Q_\lambda$ is the quality factor, $\omega_\lambda$ is the
frequency, and $(R/Q)_\lambda$ is the impedance in units of $\Omega$
for the linac definition (2 times the circuit definition).  As shown in
\cite{hoff04}, $|w_\lambda(\omega)|$ is especially large when $\omega$
is close to $\omega_\lambda$.

When motion in only one dimension is considered as in \cite{hoff04},
one obtains
\begin{equation}
I_0 \approx \frac{c}{et_b}
\frac{e^{-i\omega t_r}}{T_{12}w_\lambda}\ .
\label{eq:old1}
\end{equation}
For simplicity we again use $\mathcal{K}_\lambda =
t_b\frac{e\omega_\lambda^2}{2c^2}\roq_\lambda$ and
$\epsilon_\lambda=\frac{\omega_\lambda t_b}{2Q_\lambda}$.  For
$n_r\epsilon_\lambda\ll 1$ this leads to the approximation
\begin{eqnarray}
I_{th} &\approx&
\left\{\begin{array}{l}
- \frac{2\epsilon_\lambda}{\mathcal{K}_\lambda}
\frac{1}{T_{12}\sin(\omega_\lambda t_r)}\ {\rm for}\
T_{12}\sin(\omega_\lambda)<0,\\
\frac{2}{\mathcal{K}_\lambda|T_{12}|}
\sqrt{\epsilon_\lambda^2+(\frac{\rm{mod}(\omega_\lambda t_r,\pi)}{n_r})^2}\
{\rm else.}
\end{array}\right.
\label{eq:oldl2}
\end{eqnarray}
For $\epsilon_\lambda\ll 1$ but $n_r\epsilon_\lambda\gg 1$ the approximation
derived in \cite{hoff04} is
\begin{equation}
I_{th} \approx \frac{2\epsilon_\lambda}{\mathcal{K}_\lambda|T_{12}|}\ .
\label{eq:oldg2}
\end{equation}

Exactly the same approximation and derivation therefore leads leads
from Eq.~(\ref{eq:casea}) to
\begin{eqnarray}
I_{th\pm} &\approx&
\left\{\begin{array}{l}
- \frac{2\epsilon_\lambda}{\mathcal{K}_\lambda}
\frac{1}{T_\pm\sin(\omega_\lambda t_r+\vartheta_\pm)}\ {\rm if\ it\ is}\
>0, \\
\frac{2}{\mathcal{K}_\lambda T_\pm}
\sqrt{\epsilon_\lambda^2
+(\frac{\rm{mod}(\omega_\lambda t_r+\vartheta_\pm,\pi)}{n_r})^2}\
{\rm else.}
\end{array}\right.
\label{eq:newcl2}
\end{eqnarray}
For $\epsilon_\lambda\ll 1$ but $n_r\epsilon_\lambda\gg 1$ the approximation
derived in \cite{hoff04} is
\begin{equation}
I_{th\pm} \approx \frac{2\epsilon_\lambda}{\mathcal{K}_\lambda T_\pm}\ .
\label{eq:newcg2}
\end{equation}
The threshold current is given by $I_{th}=\min_{\pm}(I_{th\pm})$.

To clarify the here considered case, we use typical parameters for the
two HOMs: $Q_1=Q_2=10^4$, $\roq_1=\roq_2=100\Omega$,
$\omega_1=\omega_2=2\pi\cdot 2.2$GHz, $t_b=1/1.3$GHz,
$n_r-\delta=5.5$. For a decoupled optics with
$T_{12}=-10^{-6}\frac{m}{eV/c}$, $T_{14}=T_{32}=0$ we obtain a
threshold current of $I_{th}=46.40$mA that agrees to all specified
digits when computed by particle tracking and by the approximation in
Eq.~(\ref{eq:newcg2}). For a very much coupled beam transport
(abbreviation $x = -10^{-6}\frac{m}{eV/c}$ is used below) the
following threshold current is obtained:
\begin{equation}
{\mat T}=\frac{1}{\sqrt{2}}\left(\begin{array}{cccc}
 1&  x&1&3x\\
 0&  1&0&1 \\
-1&-2x&1&4x\\
 0&-1 &0&1\end{array}\right)\ \Longrightarrow\ I_{th}=20.28{\rm mA}\ .
\end{equation}
Again the threshold current computed by particle tracking and by
Eq.~(\ref{eq:newcl2}) agree to all specified digits.

\subsubsection{Small coupling} 
To cases with $|D_{11}w_1-D_{22}w_2|^2 \gg |w_1w_2 D_{12}D_{21}|$ we
refer to it as small coupling. This denomination is motivated by the
fact that when one mode is polarized in $x$ and one in $y$ direction,
this case occurs when the optical coupling is small, as can be seen in
Eq.~(\ref{eq:d21}).  Equation (\ref{eq:sqrt}) simplifies to
\begin{equation}
I_0 = \frac{e}{ct_b}
\frac{e^{-i\omega t_r}}{D_{\lambda\lambda} w_\lambda}\ ,
\label{eq:ucoupdisp}
\end{equation}
with $\lambda$ being either $1$ or $2$.

The approximation that leads from Eq.~(\ref{eq:old1}) to
Eq.~(\ref{eq:oldl2}) can again be used and it leads to
\begin{equation}
I_{th} \approx - \frac{2\epsilon_\lambda}{\mathcal{K}_\lambda}
\frac{1}{D_{\lambda\lambda}\sin(\omega_\lambda t_r)}\ {\rm for}\
D_{\lambda\lambda}\sin(\omega_\lambda t_r)<0\ ,
\label{eq:weakcoup}
\end{equation}
and similarly an approximation that corresponds to
Eq.~(\ref{eq:oldg2}) is valid. This formula has been used to argue
that an optics with very small $|D_{\lambda\lambda}|$ could be built
with extremely large threshold current. But this equation does not
apply when the $|D_{\lambda\lambda}|$ are too small, since then the
following case has to be considered.

\subsubsection{Strong coupling} 
To the case with $|D_{11}w_1-D_{22}w_2|^2 \ll |w_1w_2 D_{12}D_{21}|$
we refer to as strong coupling.  This case is especially relevant when
the mentioned coupling techniques have been used to make $D_{11}$ and
$D_{22}$ very small. One obtains
\begin{equation}
I_0^2 \approx (\frac{c}{et_b})^2
\frac{e^{-i\omega 2t_r}}{w_1w_2 D_{12}D_{21}}\ .
\label{eq:coupdisp}
\end{equation}
At the threshold current only one of the functions in the denominator
will be very large and we call this $w_\mu$. The other mode will be
indexed by $\nu$. 

In \cite{hoff04} a first order approximation in $\Delta\omega
t_b=(\omega-\omega_\mu)t_b$ and $\epsilon_\mu$ is used to obtain
Eqs.~(\ref{eq:oldl2}) and (\ref{eq:oldg2}).  Here we again expand to
first order in these quantities, which is simple since $w_\mu^{-1}$ is
linear in them so that $w_\nu^{-1}$ can be evaluated at
$\Delta\omega=0$ and $\epsilon_\nu=0$, leading to
\begin{equation}
I_0^2 \approx \frac{c}{et_b}\frac{e^{-i(\omega 2t_r+\vartheta)}}{T w_1}\ .
\end{equation}
With $Te^{i\vartheta}=\frac{et_b}{c}w_\nu(\omega_\mu) D_{12}D_{21}$,
$T\in I\!\! R^+$. Note that the exponent contains $2t_r$ instead of
$t_r$. The approximations that correspond directly to
Eqs.~(\ref{eq:newcl2}) and (\ref{eq:newcg2}) can again be applied.

In \cite{hoff04} $w_\lambda$ is evaluated for the long-range
wakefield, leading to
\begin{eqnarray}
w_\lambda &=& \roq_\lambda\frac{\omega_\lambda^2}{4c}\\
&\times&
\frac{
e^{i\omega^+ (\delta-1)t_b}\sin (\omega_\lambda  \delta t_b) -
e^{i \delta   \omega^+ t_b}\sin(\omega_\lambda [\delta-1] t_b)}
{\cos(\omega^+ t_b) -\cos(\omega_\lambda t_b)}\ .\nonumber
\end{eqnarray}
Note that here $w_\lambda$ is defined without a phase factor
$e^{-i\omega\delta t_b}$ compared to \cite{hoff04} to simplify the
notation. For an ERL, where $\delta=1/2$, one has
\begin{equation}
w_\lambda = \roq_\lambda\frac{\omega_\lambda^2}{2c}
\frac{
\cos(\omega^+\frac{t_b}{2})\sin(\omega_\lambda\frac{t_b}{2})}
{\cos \omega^+ t_b -\cos \omega_\lambda t_b}\ ,
\end{equation}
with $\omega^+t_b = \omega t_b + i\epsilon_\lambda$.
Then $\omega=\omega_\mu+\Delta\omega$ with the small quantity
$\Delta\omega t_b$. We assume that also
$\epsilon_\lambda=\frac{\omega_\lambda t_b}{2Q_\lambda}$ is small,
which is usually the case whenever BBU is relevant. A first order
expansion in these small quantities leads to
\begin{equation}
I_0^2 \approx \frac{2e^{-i \omega 2 t_r}(\Delta \omega t_b +
i\epsilon_\mu)}{\mathcal{K}_\mu \mathcal{K}_\nu D_{12}D_{21}}
\frac{\cos(\omega_\mu t_b) -\cos(\omega_\nu t_b)}
{\cos(\omega_\mu\frac{t_b}{2})\sin(\omega_\nu\frac{t_b}{2})}\ .
\label{eq:cosdif}
\end{equation}
If $I_0$ is the threshold current $I_{th}$, the right hand side has to
be a real number, requiring $\Delta\omega t_b\sin(\omega
2t_r)=\epsilon_1\cos(\omega 2t_r)$. This leads to
\begin{equation}
I_{th\mu}^2 \approx 
\frac{2\epsilon_\mu}{\mathcal{K}_\mu\mathcal{K}_\nu}
\frac{\cos\omega_\mu t_b -\cos \omega_\nu t_b}
{D_{12}D_{21}\cos(\omega_\mu\frac{t_b}{2})
\sin(\omega_\nu\frac{t_b}{2})\sin(\omega_\mu 2t_r)}\ ,
\label{eq:strongcoup}
\end{equation}
whenever this term is positive.  Whenever it is negative, the
following approximation follows from \cite{hoff04},
\begin{eqnarray}
I_{th\mu}^2 &\approx&
\frac{2}
{\mathcal{K}_\mu\mathcal{K}_\nu}
\left|
\frac{\cos\omega_\mu t_b -\cos \omega_\nu t_b}
{D_{12}D_{21}
\cos(\omega_\mu\frac{t_b}{2})\sin(\omega_\nu\frac{t_b}{2})}\right|\\
&\times&
\sqrt{\epsilon_\mu^2+(\frac{\mod(\omega_\mu 2t_r,\pi)}{2n_r})^2}\ .\nonumber
\end{eqnarray}
For $\epsilon_\mu\ll 1$ but $n_r\epsilon_\mu\gg 1$ one obtains
\begin{equation}
I_{th\mu}^2 \approx
\frac{2\epsilon_\mu}{\mathcal{K}_\mu\mathcal{K}_\nu}
\left|
\frac{\cos\omega_\mu t_b -\cos \omega_\nu t_b}
{D_{12}D_{21}\cos(\omega_\mu\frac{t_b}{2})\sin(\omega_\nu\frac{t_b}{2})}
\right|\ .
\end{equation}

These formulas are to be evaluated for $\mu=1$, $\nu=2$ and for
$\mu=2$, $\nu=1$, and the smaller of the two resulting currents is
the threshold current,
\begin{equation}
I_{th} = \min_{\mu\in\{1,2\}}\{I_{th\mu}\} \ .
\end{equation}

An interesting observation is that for two modes with similar
$Q_\lambda$, $\roq_\lambda$, and $\omega_\lambda$, the ratio of the
threshold current with and without coupled optics can be found by
comparing Eqs.~(\ref{eq:strongcoup}) and (\ref{eq:weakcoup}) and it is
proportional to $\sqrt{1/\epsilon_\lambda}\propto
\sqrt{Q_\lambda}$. For cavities that are optimized for large currents
by means of sophisticated HOM damping, the advantage of a coupled
optics therefore decreases. This effect is independent of the length
of the return loop and is already relevant for a single cavity.

\begin{figure}
\centering
\includegraphics[width=\linewidth,clip]{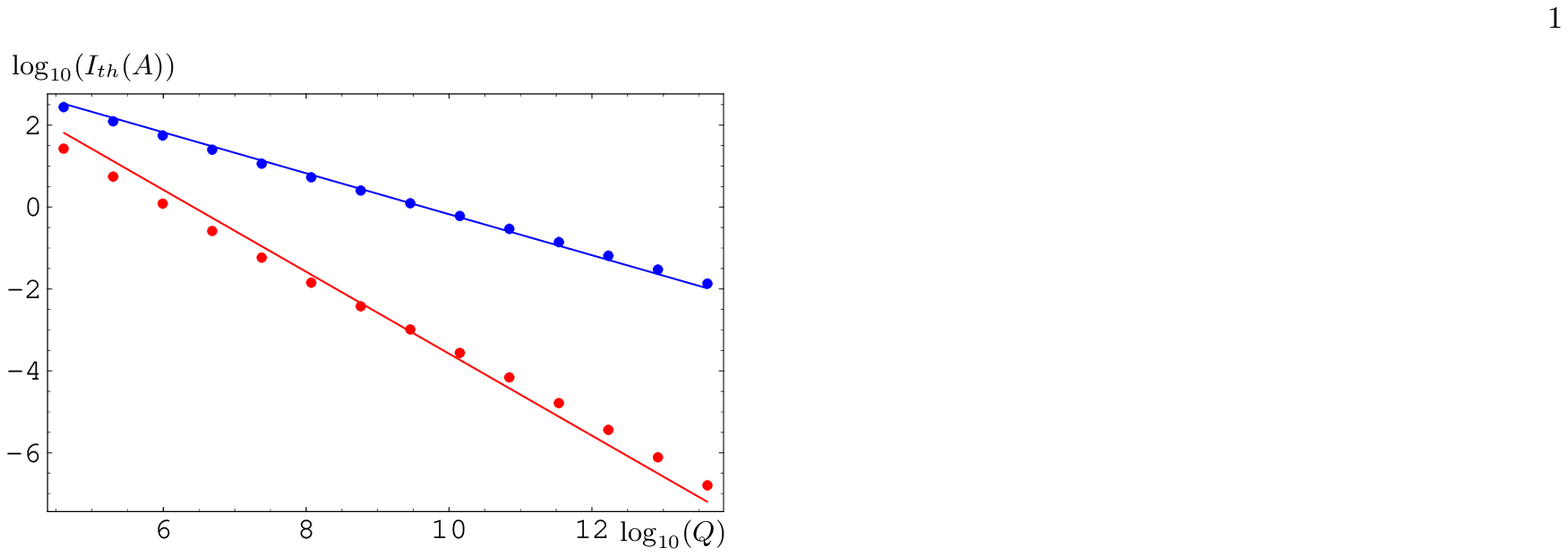}
\caption{The threshold with two modes in one cavity scales with
$Q^{-1}$ for a decoupled and with $Q^{-\frac{1}{2}}$ for a fully
coupled optics. The advantage of coupling thus diminishes for low Q.}
\label{fg:qcomp}
\end{figure}

Figure \ref{fg:qcomp} refers to a linac with one cavity and
$t_r=555.5t_b$ that has two HOMs, one with $\omega_1=2.2$GHz polarized
in $x$ and one with $\omega_2=2.3$GHz polarized in $y$ direction,
$\roq=100\Omega$, and $Q_x=Q_y$ is varied. The bottom data refer to a
decoupled optics and the top data to a fully coupled optics with
$T_{12}=0$ and $T_{34}=0$. A double-logarithmic plot is shown, which
makes it apparent that the threshold current for a decoupled optics
decreases with $Q^{-1}$, as indicated by the line with slope $-1$. A
closer look shows that the slope is only accurately $-1$ for
relatively small and relatively large $Q$ where either approximation
\ref{eq:oldl2} or \ref{eq:oldg2} hold. A totally coupled optics with
$T_{12}=0$ and $T_{34}=0$ leads to a larger threshold current than
without coupling, but when $Q$ of the modes is reduced, the threshold
current increases only with $Q^{-\frac{1}{2}}$ as indicated by a line
with slope $-\frac{1}{2}$.  The advantage of coupling decreases
proportionally to $\sqrt{Q}$.

\begin{figure}
\centering
\includegraphics[width=\linewidth,clip]{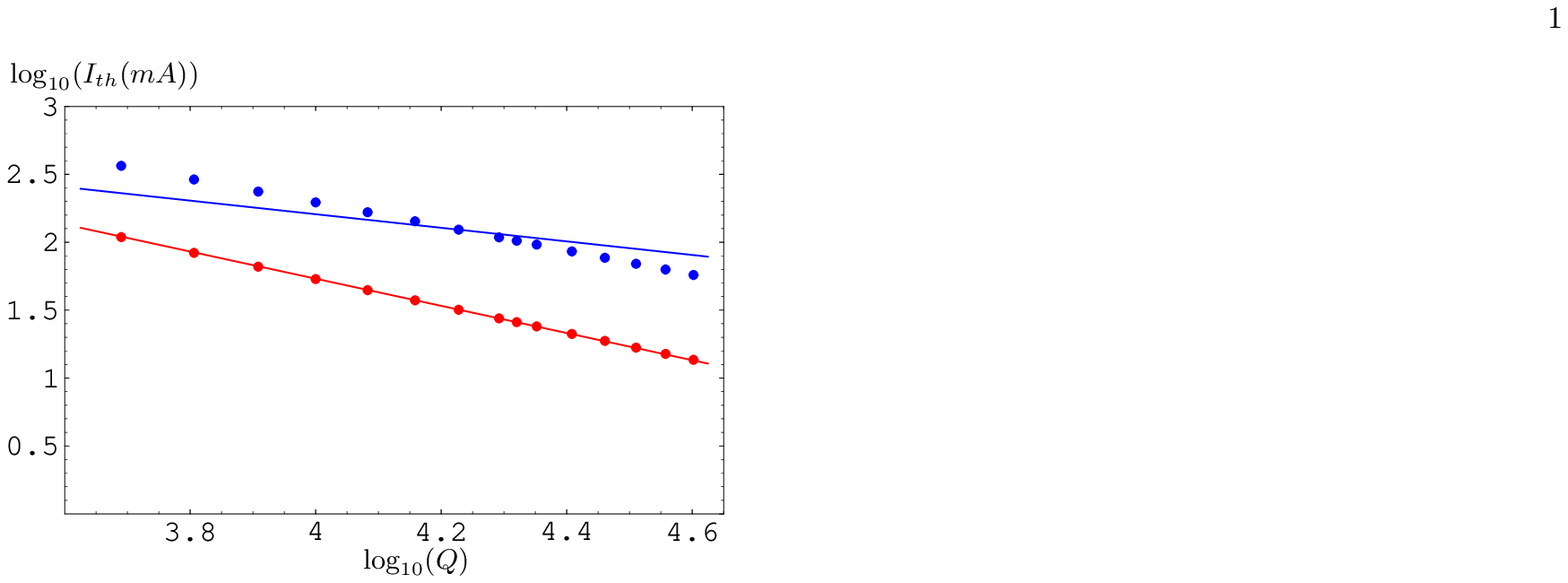}
\caption{The threshold increase due to coupling of the x-ray ERL
upgrade of CESR scales roughly with $\sqrt{Q}$.}
\label{fg:qerlcomp}
\end{figure}

Figure \ref{fg:qerlcomp} similarly shows the advantage of polarizing
the higher order modes in the ERL that is proposed to upgrade the CESR
ring at Cornell University \cite{erlpac05}. When the HOMs are
polarized in $x$ and $y$ direction and the optics is completely
coupled by $T_{12}=0$ and $T_{34}=0$, the threshold current is larger
than without coupling, but again the advantage is smaller when the
HOMs are damped more strongly by HOM absorbers. In fact, for the HOMs
that are computed for $7$-cell cavities \cite{Liepe03}, the increase
of the threshold current due to coupling is only a factor of $2.5$ as
can be seen in Tab~\ref{tb:cesr} when the HOM frequencies in $x$ and
$y$ are separated by 10MHz. However, the figure shows that when many
cavities are present, as in the ERL where there are 320, the scaling
is not as simple as in the case of a single cavity and the advantage
of coupling does not decrease as strongly with decreasing $Q$.

For Tab.~\ref{tb:cesr} the 320 cavities of the ERL upgrade of CESR had
nominal HOM frequencies of $f_x=1.87394$GHz with horizontal
polarization. The mode with vertical polarization is $f_y=f_x-\Delta
f_{xy}$. The cavities have HOM frequencies that have a Gaussian
distribution around these values with rms width $\sigma_{RF}$. We used
500 different random distributions of the frequencies and display the
average threshold current $I_{th}$ as well as the rms $\sigma_I$ of
the 500 resulting thresholds. All modes have $\roq=109.60\Omega$ and
$Q=20912.4$.

\begin{table}
\caption{Threshold currents for the two most significant HOMs of the
Cornell ERL.\\}
\label{tb:cesr}
\begin{tabular}{|ccccc|}
\hline
$\Delta f_{xy}$ & Coupling & $\sigma_{RF}$ & $I_{th}$ & $\sigma_I$\\
\hline
10MHz & NO  &   0MHz & 25.8mA &   0mA \\
10MHz & YES &   0MHz &  103mA &   0mA \\
10MHz & NO  & 1.3MHz &  280mA &  43mA \\
10MHz & YES & 1.3MHz &  867mA & 100mA \\
60MHz & NO  &  10MHz &  418mA &  69mA \\
60MHz & YES &  10MHz & 2420mA & 433mA \\
\hline
\end{tabular}
\end{table}

\subsection{Comments about numerical solutions}
\begin{figure}
\centering
\includegraphics[width=\linewidth,clip]{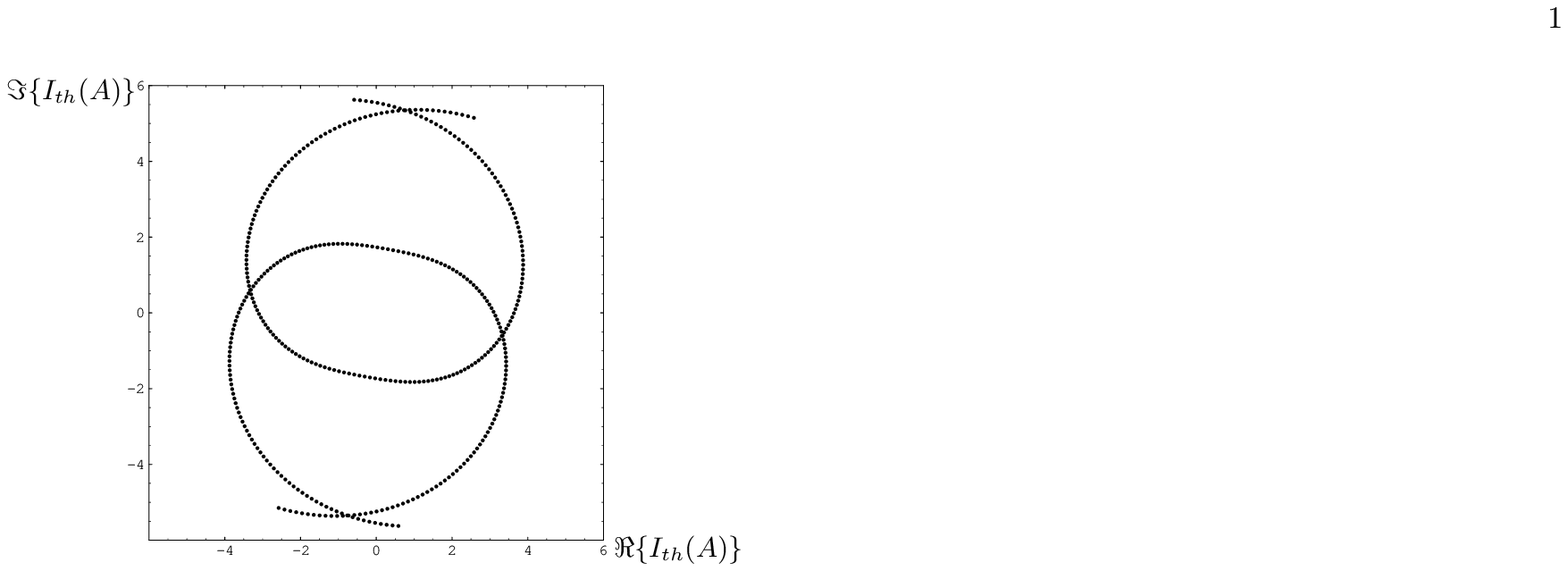}
\caption{Two eigenvalues trace out loops for $I_0$ in the complex
plain while $\omega$ is varied.}
\label{fg:loops}
\end{figure}

When the threshold current for two polarized modes at $\omega_\nu$ and
$\omega_\mu$ should be found by solving Eqs.~(\ref{eq:ucoupdisp}) and
(\ref{eq:coupdisp}) numerically, the eigenvalues are plotted in the
complex plain, and the intersections with the real axis are sought
that lead to the smallest current. An example of the two eigenvalues
plotted in the complex plain is shown in Fig.~\ref{fg:loops}. The
eigenvalues are largest in the vicinity of $\omega=|{\rm
mod_\pm}(\omega_\lambda,\frac{2\pi}{t_b})|$, $\lambda\in\{1,2\}$,
since there either $w_\mu$ or $w_\nu$ become very large.  The
subscript on the mod function indicates that ${\rm
mod_+}(x,1)\in[0,1]$ and ${\rm
mod_-}(x,1)\in[-\frac{1}{2},\frac{1}{2}]$.

Furthermore, the eigenvalues trace out loops around the origin of the
complex plain about once per $\frac{2\pi}{t_b}$ variation of $\omega$,
due to the exponential factor in Eq.~(\ref{eq:ucoupdisp}). We
therefore vary $\omega$ only in a $\pm \frac{2\pi}{t_b}$ interval
around each HOM frequency. This speeds up the search for eigenvalues
by a factor proportional to $n_r$, which can be very large.

\begin{figure}
\centering
\includegraphics[width=\linewidth,clip]{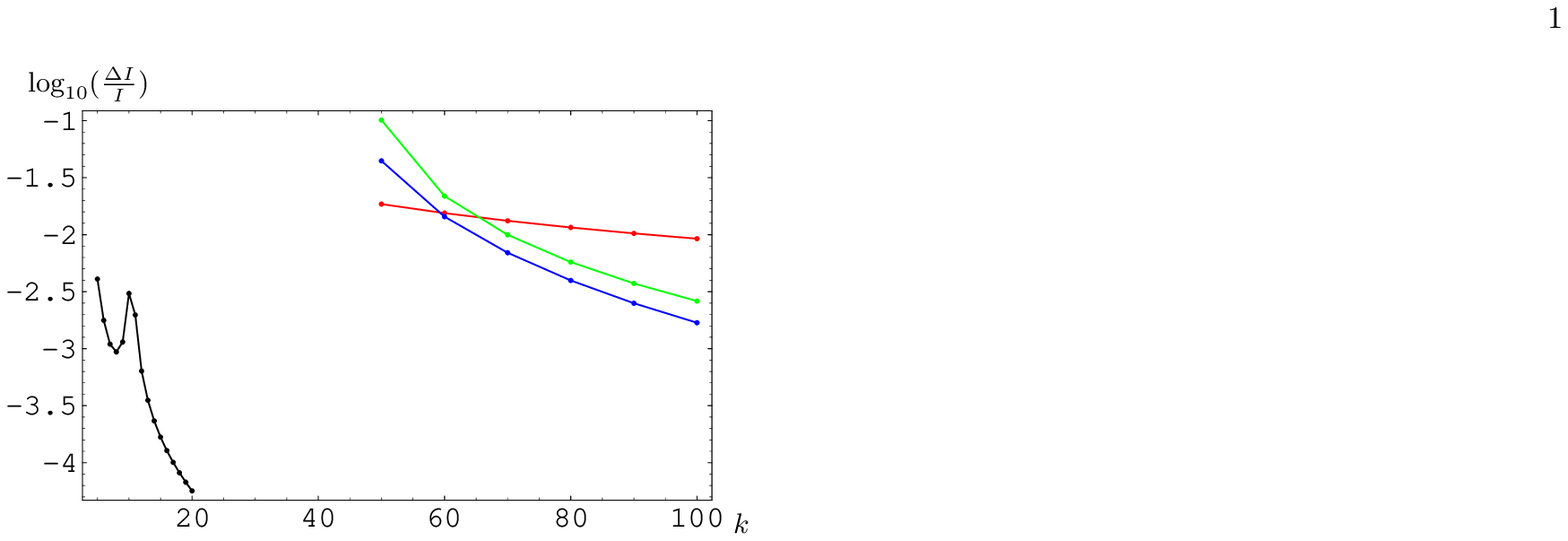}
\caption{Variation of the relative accuracy $\frac{\Delta I}{I}$ for
scanning $\omega-\omega_\mu\in[-\frac{2\pi}{t_r},\frac{2\pi}{t_r}]$ by
$k$ points. By far the best accuracy is achieved with the elliptical
extrapolation of 4 points around the real axis (black dots). For an
approximation by a line between two points (red, largest at large
$k$), a second order polynomial $y(x)$ fitted to three points (green,
second largest at large $k$), and a third order polynomial $y(x)$
fitted to four pints (blue, third largest at large $k$) all lead to
much worth accuracies and/or computation times.}
\label{fg:accuracy}
\end{figure}

A simple approach would be to plot all eigenvalues in the complex
plain and to select the smallest eigenvalue that is reasonably close
to the complex plain. Large factors in speed can be gained when the
loops that are traced out by each eigenvector can be interpolated and
their intersection with the real axis can be found, since the loops
have to be scanned much less densely. We have found that using values
of $I_0(\omega)$, where two are above and two below the real axis, and
fitting an upright ellipse to these values is a very good
parametrization. The accuracy achieve for distributing $k$ particles
in the described region around each HOM frequency lead to the accuracy
documented in Fig.~\ref{fg:accuracy}.

\subsection{Comparison of results for polarized modes and coupling}
\begin{figure}
\centering
\includegraphics[width=\linewidth,clip]{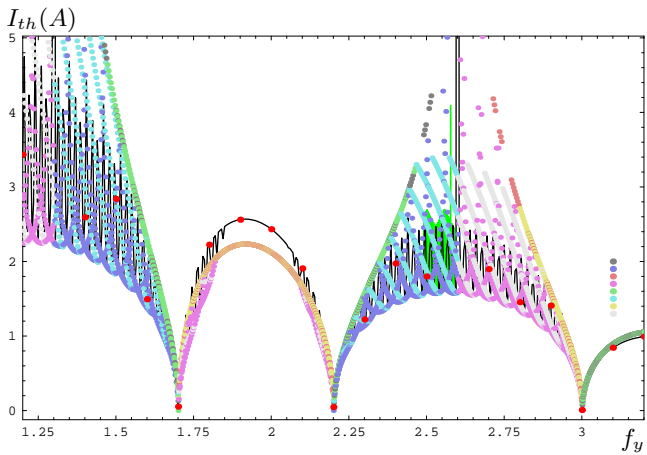}
\caption{Threshold current for one horizontal HOM with $f_x=2.2$GHz as
a function of the frequency $f_y$ of a vertical HOM. Black curve:
dispersion relation, Red dots: tracking, Colored curves: the different
applicable approximations derived above.}
\label{fg:tracking}
\end{figure}

To demonstrate the excellent agreement between these numerical
solutions and tracking, we depict Fig.~\ref{fg:tracking} where one HOM
with horizontal polarization has been fixed at $f_x=2.2$GHz, and
another has been varied for $f_y\in [1.2,3.2]$GHz. The optics was
completely coupled with $T_{12}=T_{34}=0$ and $T_{31}=-T_{14}=10^{-6}
\frac{m}{eV/c}$. Several things can be observed: (1) since tracking is
relatively time consuming, only relatively few frequencies for the
second HOM have been evaluated, but all of them lie exactly on the
curve that follows form the dispersion relation. (2) The threshold
current varies strongly when the second HOM is varied, but the
agreement with tracking shows that this is not numerical noise but a
consequence of the coupling between the two polarized modes whose
frequencies are very far apart. (3) There are frequencies where the
threshold current is relatively small, these are frequencies where
$\cos(\omega_1 t_b)\approx\cos(\omega_2t_b)$ in agreement with
Eq.~(\ref{eq:cosdif}). The displayed minima appear at $(3\cdot 1.3 -
2.2)$GHz, $2.2$GHz, and $(4\cdot 1.3 - 2.2)$GHz. But the regions with
reduced threshold are relatively wide; in particularly they are much
wider than the width of HOM resonances.

\begin{figure}
\centering
\includegraphics[width=\linewidth,clip]{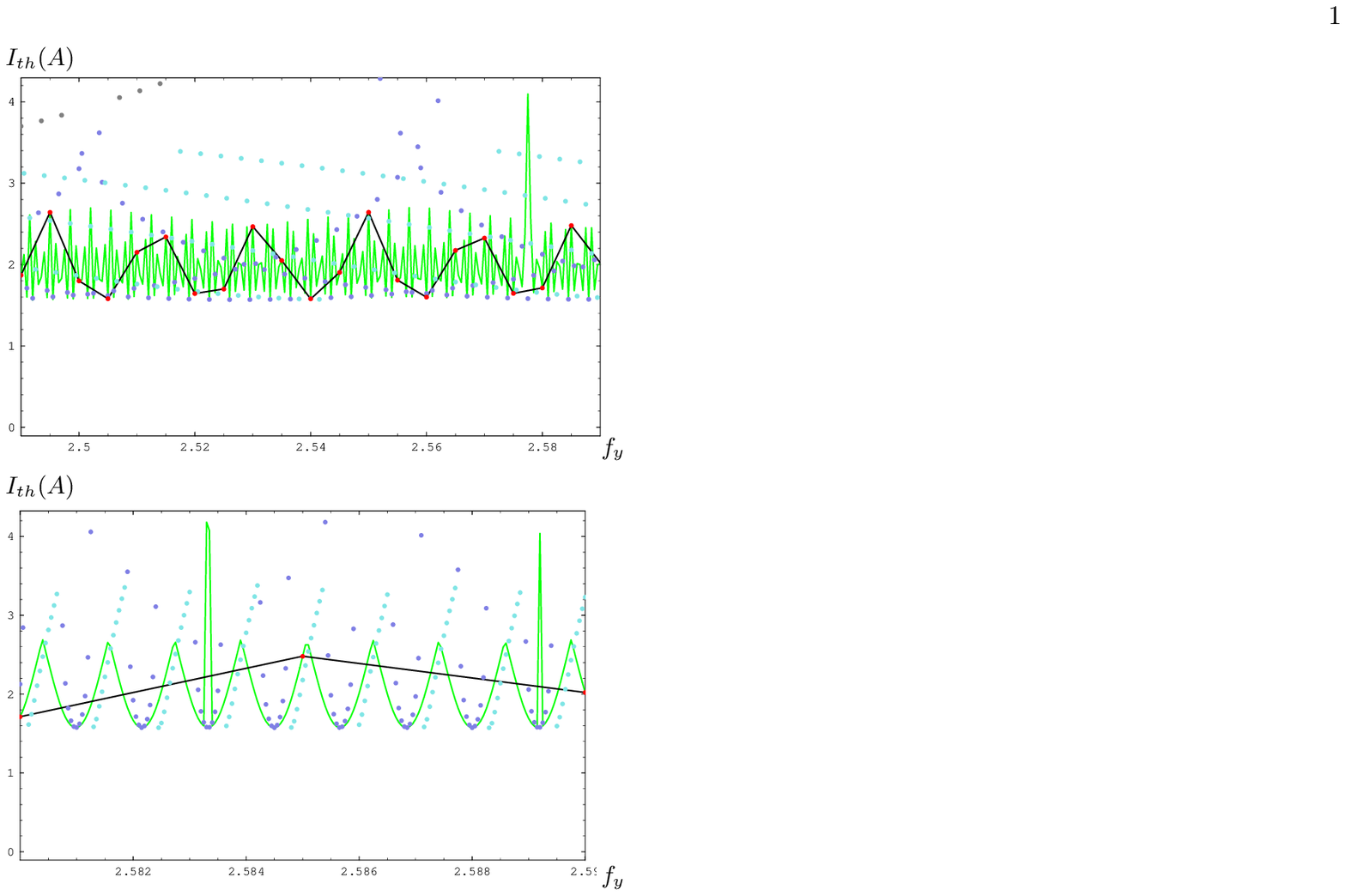}
\caption{Magnification of regions in Fig.~\ref{fg:tracking}}
\label{fg:parts}
\end{figure}

Depicted in color are the values obtained by approximations derived
above. It is apparent that the approximations are not always very
good, especially that they lead to values that are too large. A
magnification in Fig.\ref{fg:parts} shows that the reason for that is
that the approximate formulas lead to parabolic shapes that have the
correct minimum value, but not the correct width. This is important
since it shows that the formulas can be used to find the correct
minimum value as a conservative estimate. Furthermore, the
magnifications again show the good agreement with tracking results.

Here an important note is in place. Once place where a strong dip
occurs is when $\omega_1\approx\omega_2$. And this dip is much wider
than the width of the HOM resonance of $\omega_\lambda/Q\lambda$,
showing that the two modes clearly do not decouple when they are
separated by more than their width. For nominally circular symmetric
cavities, HOMs are not degenerate due to construction errors and each
mode splits into two mode with typically a few MeV distance. But the
dip is much wider than that, showing that an appropriate advantage of
BBU suppression by damping can in general only be realized when
polarized cavities are designed, i.e. cavities where the horizontal
and vertical dimensions are designed to be slightly different, leading
to HOM frequencies that differ by several 10MeV in the two planes.

\begin{figure}
\centering
\includegraphics[width=\linewidth,clip]{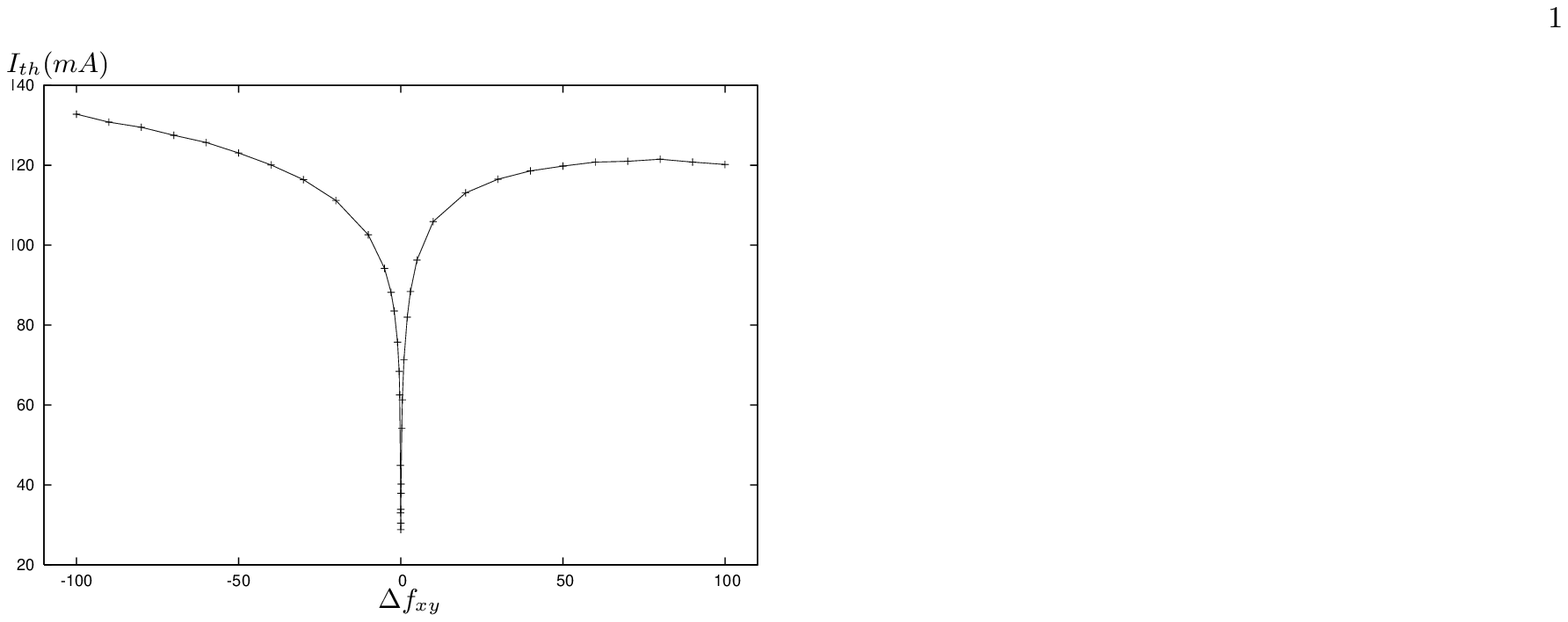}
\caption{Dependence of $I_{th}$ on $\Delta f_{xy}$ for the Cornell ERL.}
\label{fg:deltaf}
\end{figure}

The question arises how far the HOM frequencies have to be apart. In
Fig.~\ref{fg:deltaf} we show for the Cornell ERL how $I_{th}$ changes
with $\Delta f_{xy}$. In order to avoid averaging over many different
frequency distributions we have here chosen the same two HOM
frequencies for each of the 320 cavities. The data indicates that a
mode separation of $60$MHz is sufficient. This has been used to
compute the very large recirculating BBU threshold current of close to
2.5A for this accelerator. It should be noted that only the two
dominant HOMs were considered in this simulation and a detailed study
is in order when such high currents are sought. The Cornell ERL is
designed for 100mA and this calculation indicates that polarized
cavities with a coupled optics provide a very comfortable safety
margin with respect to this instability.

\subsection{Approximation for $N$ polarized HOMS in one cavity}
As a first approximation one can assume that one component of the
eigenvectors in Eq.~\ref{eq:eigen} will be very large. This would lead
to a decoupling of HOMs so that each HOM could be treated separately,
as for a single degree of freedom. One could therefore derive the
threshold with the smallest current for each individual HOM, and this
would approximate the threshold for the complete accelerator. When all
these thresholds are very large, one has to investigate the next
approximation, where the eigenvector has two dominant components. In
this case the eigenvalues are determined from a $2\times 2$ matrix
corresponding to Eq.~(\ref{eq:eigen2}). The threshold current has to be
computed for each pair of HOMs. The smallest current obtained for one
pair of polarized modes then approximates the threshold current of the
full accelerator.

\section{Polarized HOMs in many cavities and for multiple turns}
Recirculating linacs with many cavities and several recirculation
loops have been considered early on \cite{Bisognano87_01, Krafft89}.
In \cite{hoff04} a description for arbitrary recirculation times has
been presented. Here we want to extend this description to include
orbit coupling and polarized modes. As far as possible we retain the
notation of these earlier papers. The $N$ higher order modes, which
can be associated with different cavities, are numbered by an index
$i$.  The $N_p$ passes through the linac are numbered by an index $I$.
The horizontal and vertical phase space coordinates that the beam has
at time $t$ in the HOM $i$ during turn $I$ is denoted $\vec
z_i^I(t)=(x_i^I(t),p_{xi}^I(t),y_i^I(t),p_{yi}^I(t))^T$.  The $4\times
4$ transport matrix that transports the phase space vector $\vec
z_j^J$ at HOM $j$ during turn $J$ to $\vec z_i^I$ is denoted ${\mat
T}_{ij}^{4I\!J}$ and the time it takes to transport a particle from
the beginning of the first turn to HOM $i$ during turn $I$ is denoted
$t_i^I$.  The beam is propagated from after HOM $i-1$ to after HOM $i$
by
\begin{equation}
\vec z_i^I(t) = {\mat T}_{ii-1}^{4II} \cdot \vec
z_{i-1}^I(t-[t_{i}^I-t_{i-1}^I])+\frac{e}{c}\vec V_i(t)\ ,
\end{equation}
with $\vec V_i = V_i(0,\cos\theta_i,0,\sin\theta_i)^T$.  This equation
can be iterated to obtain the phase space coordinates as a function of
the HOM strength that creates the orbit oscillations.  With the matrix
\begin{equation}
{\mat T}_{ij}^{I\!J}
=
\left(
\begin{array}{cc}
({\mat T}_{ij}^{4I\!J})_{12} & ({\mat T}_{ij}^{4I\!J})_{14}\\
({\mat T}_{ij}^{4I\!J})_{32} & ({\mat T}_{ij}^{4I\!J})_{34}
\end{array}
\right)
\end{equation}
one obtains
\begin{eqnarray}
\vec x_i^I(t)
&=&
\sum_{J=1}^{I}\sum_{j=1}^{N_{IJ}(i-1)} {\mat T}_{ij}^{I\!J}\vec e_i \frac{e}{c}
V_j(t-[t_i^I-t_j^J])\ ,\label{eq:xfromv}\\
N_{I\!J}(i-1)&=&
\begin{cases}
  N, &\text{if $I\ne J$;} \\
  i-1, &\text{if $I = J$.}
\end{cases}
\end{eqnarray}

The strength $V_i(t)$ of the HOM $i$ with polarization direction $\vec
e_j$ is created by all particles that have traveled through that HOM
via the integral
\begin{equation}
V_i(t)=\int_{-\infty}^\infty \sum_{I=1}^{N_p}W_i(t-t')I_i^I(t')\vec
e_i^T\vec x_i^I(t')dt'\ ,
\label{eq:homsum}
\end{equation}
where $I_i^I(t)$ is the current at time $t$ that the fraction of the
beam has which passes the HOM $i$ on turn $I$.  Note that $W(t-t')=0$
for $t'>t$. Combining this with Eq.~(\ref{eq:xfromv}) leads to the
following integral-difference equation:
\begin{eqnarray}
V_i(t)&=&\int_{-\infty}^\infty\sum_{I=1}^{N_p} W_i(t-t')I_i^I(t')\\
&\times&\frac{e}{c} \sum_{J=1}^I\sum_{j=1}^{N_{I\!J}(i-1)}D_{ij}^{I\!J}
V_j(t'-[t_i^I-t_j^J])dt'\ ,\nonumber\\
D_{ij}^{I\!J} &=& \vec e_i^T{\mat T}_{ij}^{I\!J}\vec e_j\ .
\end{eqnarray}

This equation is identical to that obtained for one degree of freedom,
only that $({\mat T}_{ij}^{I\!J})_{12}$ is replaced by $D_{ij}^{I\!J}$.
The following treatment for obtaining the threshold current is
therefore identical to that in \cite{hoff04}. For completion it is
here presented in simplified form, and recommendations for numerical
solutions are given.

Now the approximation of short bunches is used.  The current is given
at time $t$ by pulses that are equally spaced with the distance $t_b$,
\begin{equation}
I_i^I(t)=\sum_{m=-\infty}^\infty I_0 t_b \delta(t-t_i^I-m t_b)\ .
\end{equation}
This reduces the integral to a sum,
\begin{eqnarray}
V_i(t) &=& \frac{e}{c}I_0 t_b \sum_{m=-\infty}^\infty
\sum_{I=1}^{N_p} W_i(t-t_i^I-m t_b)\label{eq:hommultsum}\\ &\times&
\sum_{J=1}^I\sum_{j=1}^{N_{I\!J}(i-1)} D_{ij}^{I\!J}V_j(m
t_b+t_j^J)\ .\nonumber
\end{eqnarray}
This leads to
\begin{eqnarray}
\tilde V^\Sigma_{i,t_i^L/t_b}(\omega)
&=&
\frac{e}{c}I_0 t^2_b \sum_{n=-\infty}^\infty
\sum_{m=\infty}^\infty
\sum_{I=1}^{N_p} W_i(m t_b+t_i^L-t_i^I)\\ &\times&
\sum_{J=1}^I\sum_{j=1}^{N_{I\!J}(i-1)} D_{ij}^{I\!J}V_j([n-m]
t_b+t_j^J)e^{i\omega nt_b}\ .\nonumber\\
&=&
\frac{e}{c}I_0 
\sum_{I=1}^{N_p} \tilde W^\Sigma_{i,t_i^L-t_i^I}
\sum_{J=1}^I\sum_{j=1}^{N_{I\!J}(i-1)} D_{ij}^{I\!J}
\tilde V^\Sigma_{j,t_j^J/t_b}(\omega)
\ .\nonumber
\end{eqnarray}

If a vector $\vec V$ is introduced that has the coefficients $\tilde
V^\Sigma_{i,t_i^I}$, this equation can be written in matrix form,
\begin{equation}
\frac{1}{I_0}\vec V={\mat M}(\omega)\vec V\ ,
\label{eq:vmat}
\end{equation}
with the matrix coefficients
\begin{eqnarray}
M_{ij}^{LJ} &=& \frac{e}{c}t_b\sum_{I=J+\Theta_{j,i}}^{N_p}\tilde
W_{i,[t_i^L-t_i^J]/t_b}(\omega)D_{ij}^{I\!J}\nonumber\\ \Theta_{j,i}&=&
\begin{cases}
  1, &\text{if $j \ge i$;} \\
  0, &\text{otherwise.}
\end{cases}
\label{eq:mmat}
\end{eqnarray}
Note that $\Tilde W^\Sigma_{i,[t_i^L-t_i^I]/t_b} = \Tilde
W^\Sigma_{i,\delta_i(I,L)}e^{i\omega Top(\frac{t_i^I-t_i^L}{t_b})t_b}$
where ${\rm Top}(x)$ is the smallest integer that is equal to or
larger than $x$ and $\delta_i(I,L) = {\rm mod}(t_i^I-t_i^L,t_b)$.
With Kronecker $\hat\delta_{ik}$ this determines the matrices ${\mat
W}$ and ${\mat U}$ to be
\begin{eqnarray}
W_{ik}^{LI} &=&\frac{e}{c}t_b w_i(\delta(I,L)) e^{i\omega{\rm
Top}(\frac{t^I-t^L}{t_b})t_b} \delta_{ik}\ ,\label{eq:wdef}\\
U_{kj}^{I\!J} &=&T_{kj}^{I\!J}\Theta_{I,J+\Theta_{j,k}}\
.\label{eq:udef}
\end{eqnarray}

For each frequency $\omega$, $I_0^{-1}$ is an eigenvalue of ${\mat
M}(\omega)$.  Since the eigenvalues are in general complex, but $I_0$
has to be real, the threshold current is determined by the largest
real eigenvalue of ${\mat M}(\omega)$.  The matrix has the properties
\begin{equation}
{\mat M}(\omega+\frac{2\pi}{t_b})={\mat M}(\omega)\ ,\ \ {\mat
M}(-\omega^*)={\mat M}^*(\omega)\ ,
\end{equation}
and it is therefore again sufficient to investigate $\omega\in[0,\pi
/t_b]$ to find the threshold current.

Note that $V_N^{N_p}$ never appears since the last kick on the last
turn does not feed back to any HOM, so that the dimension of ${\mat
M}$ can be reduced by one to $N\cdot N_p-1$.  Furthermore the
dimension can be reduced when two fractional parts $\delta_i^I$ and
$\delta_i^J$ are equal since then $V_i^I$ and $V_i^J$ are identical.

\subsection{Multi turn operation and cavity misalignments}
Since the formalism presented here that includes polarized modes and
coupled optics is identical to the formalism in one degree of freedom,
only that $({\mat T}_{ij}^{I\!J})_{12}$ has been replaced by
$D_{ij}^{I\!J}$, all conclusions about multi turn recirculating and
multi turn ERLs hold. For example, in \cite{hoff04} it was concluded
that for one HOM and for $N_p$ passes through the linac, the threshold
current should roughly scale as $N_p(2N_p-1)$. The origin of this
conclusion results from the double sum
$\sum_{I=1}^{N_p}\sum_{J=1}^{I}$. Since the same summation appears,
this conclusion holds also for polarized modes with coupling.

For misaligned cavities, HOMs are excited even when the current is
smaller than the threshold current. This can lead to large beam
excursions, and in \cite{hoff04} it was analyzed for what currents
these excursions become extremely large. It was found that the BBU
threshold current is always smaller than the current for which these
orbit excursions would get very large. Since the here presented
formalism with coupling and polarized modes has the same formal
structure, this conclusion again holds.

\subsection{Comments about numerical solutions}
As pointed out above where numerical solutions for one return loop and
two HOMs were found, it is very essential to systematically search for
real values for the eigenvalues of ${\mat M}$. Each eigenvalue traces
out curves in the complex plain when $\omega$ is varied in the region
$[0,\pi/t_b)$, however eigenvalue finders usually do not return
eigenvalues in any particular order, so that these curves cannot be
observed easily. If they could be observed, then ellipses could be
fitted to these curves and the intersection of the curve with the real
axis could very efficiently be found for each eigenvector.

We therefore recommend a sorting algorithm that sorts eigenvalues
rather robustly: (1) Normalize each eigenvector. (2) Sort these
vectors according to their largest component, i.e. the vector which
has its largest component in position 1 is the first vector, if there
are more than one of this kind, the one with the largest coefficient
can be chosen as first vector, etc. (3) Associate the eigenvalues in
the order of these eigenvectors. Small changes of $\omega$ do not
change the relative size of the eigenvector elements much. The
intersection of the curve $\lambda_i(\omega)$ with the real axis can
now be found for each eigenvector. This procedure leads to an
enormous speed advantage over simply scanning all eigenvalues for a
mesh of $\omega\in[0,\pi/t_b]$, and choosing the largest eigenvalue
that is reasonably close to the real axis.

\end{document}